\documentclass[nohyper,12pt,letterpaper]{JHEP}
\usepackage{epsfig}

\newfont{\frak}{eufm10 scaled 1200}

\newfont{\Bbb}{msbm10 scaled 1200}     
\newcommand{\mathbb}[1]{\mbox{\Bbb #1}}
\DeclareSymbolFont{AMSa}{U}{msa}{m}{n}
\DeclareSymbolFont{AMSb}{U}{msb}{m}{n}
\let\Box\relax
\DeclareMathSymbol{\Box}{\mathord}{AMSa}{"03}

\def \eqn#1#2{\begin{equation}#2\label{#1}\end{equation}}

\title{An upper bound on the number of e-foldings}

\author{T. Banks \\
  Department of Physics and Astronomy - NHETC\\
  Piscataway, NJ 08540\\
  and\\
  Department of Physics, SCIPP\\
  University of California, Santa Cruz, CA 95064\\
E-mail: \email{banks@scipp.ucsc.edu}}

\author{ W. Fischler,  \\
    Department of Physics\\
    University of Texas, Austin, TX 78712\\
E-mail: \email{fischler@physics.utexas.edu}}

\abstract{If the present acceleration of the universe is due to an
asymptotically de Sitter universe with small cosmological
constant, and the principle of Cosmological Complementarity is
valid, then the number of e-foldings during inflation is bounded.
}


\preprint{\hepth{} \\RUNHETC-2003-47 \\SCIPP-03/29\\
UTTG-04-03\\ NSF-KITP-04-28}
\begin{document}


\section{Introduction}

A remarkable series of  observations
\cite{deBernardis:2001xk}-\cite{Riess:1998cb} ( for a review see
\cite{Primack:2002th}), indicate that we live in an accelerating
universe. It is quite tempting to speculate that the source of
this acceleration is a small positive cosmological constant,
leading to an asymptotically de Sitter (dS) space. If this is
indeed the case, there is a restriction on the number of
e-foldings during inflation.

We emphasize that when we say the observed acceleration is due to
a cosmological constant, we mean that space-time is asymptotically
de Sitter, with the value of the cosmological constant that fits
the data. Our bound would not apply to a model of a meta-stable dS
minimum, which decayed into an asymptotic space-time with
vanishing cosmological constant. More generally, it is important
to realize that theories of quantum gravity are defined with fixed
asymptotic conditions.   The same effective Lagrangian may have
solutions which describe the physics of completely different
quantum models of gravity. In the second half of this paper we
will argue that inflation models with large numbers of e-foldings
and a positive ``true" cosmological constant, are not
asymptotically dS, and have only a temporary dS phase.   Their
true asymptotic behavior depends on issues in quantum gravity
which are not yet resolved. In a brief comment at the conclusion
of this paper, we will take up the difficult problem of how to
differentiate a truly asymptotically dS space-time from such a
meta-stable scenario, by current observations.

The existence of a small cosmological constant,
$\lambda$\footnote{When equations do not appear dimensionally
homogeneous, the reader should understand that we are using Planck
units.}, makes the universe eventually appear to a local observer
as a finite cavity of size ${\lambda}^{-1/2}$ \cite{lennybf}.
Second, as was shown in a recent paper \cite{flp}, this finite
size cavity can only accommodate a limited amount of entropy
stored in field theoretical degrees of freedom. It was shown in
\cite{flp} that this limited amount of entropy scales like
${\lambda}^{-3/4}$. Any excess entropy beyond this bound has to be
encoded into black holes or imprinted onto the walls of the
cavity. That excess entropy in turn is limited to be smaller than
the entropy of empty de Sitter space \cite{GH,rbousso}.

This third version of our paper is written to clarify the basic
assumption which underlies our bound.   It was left implicit in
the first version and explained only cursorily in the second.  We
have found that many people have been confused about our claim,
because they did not understand this basic assumption.

In \cite{mcosmo} we announced a principle we called {\it
Cosmological Complementarity}.  According to this principle, the
physics of an asymptotically de Sitter space-time is completely
described by a pure state in a Hilbert space of finite dimension,
determined by the value of the cosmological constant.   This
principle cannot be derived from the covariant entropy
bound \cite{raph} or any simple generalization of it.   It was
meant to parallel the notion of Black Hole
Complementarity\cite{tHsuss}, which is necessary to the unitary
resolution of the black hole information problem.  Complementarity
for dS space does not however follow from Complementarity for
black holes.   Cosmological Complementarity was meant to be a
guide in formulating a quantum theory of asymptotically dS
space-times.   It is the basic assumption on which our bound on
the number of e-foldings of inflation rests.

The other important assumption we will make is that at early
times, the standard description of the inflationary universe in
terms of quantum field theory in curved space-time makes sense
over the entire inflationary patch.   There are several reasons
for this.  One is that there are no obvious inconsistencies in the
field theoretic treatment.  Indeed, if the true cosmological
constant were zero, inflation theorists would use the predictions
for fluctuations on the scale of this patch to try to fit the data
that would be seen as more and more of the CMB fluctuations came
into our horizon.   However the most important reason for making
this assumption is that it affects the state of an observer inside
the cosmological horizon volume.   The Bunch Davies (BD) wave
function is typically written in terms of momentum modes in the
flat coordinate system for the expanding half of the approximate
dS space of the inflationary epoch. It contains no correlations
between different modes. However, if we rewrite it in terms of
wave functions concentrated in the cosmological horizon volume of
a given observer in the asymptotically dS space, and wave
functions with support outside that volume, then there are
correlations between the inside and outside modes.   Thus, the
standard inflationary picture contains correlations between the
observables in a single cosmological horizon volume, and those of
a potentially larger set of degrees of freedom.   Indeed, one of
the prime virtues of inflation is its ability to explain
correlations between widely separated objects that might otherwise
appear causally disconnected.

We will count the entropy implied by the inflationary picture,
with a variety of assumptions about what the products of reheating
are.   We argue that if it is larger than the dS entropy of
asymptotic dS space, that the density matrix on the finite
dimensional dS Hilbert space cannot be pure.  This contradicts the
assumption of Cosmological Complementarity.

The reader is perfectly justified in concluding that he can avoid
having to think about a bound on the number of e-foldings by
rejecting our assumption about purity of the density matrix. This
is an assumption which is virtually impossible to check in any
experimental way. We remind the reader who is inclined to do this,
that similar remarks could have been made about unitarity of the
S-matrix for black hole production and decay in asymptotically
flat space-time.   It is only because examples exist of exactly
unitary theories of quantum gravity with asymptotically flat and
AdS geometries that we have strong arguments for a unitary
S-matrix in this case\footnote{The arguments are not so strong as
to have convinced everyone in the field.}. Once we allow for the
possibility of violation of unitarity for a single causally
connected patch in any space-time we must ask ourselves why it
doesn't occur in general.

Indeed, models with a very large number of e-foldings often appear
to lead to the phenomenon of eternal or self-reproducing
inflation, when analyzed by the methods of quantum field theory in
curved space-time.   Such models appear to violate unitarity for a
single observer, even when there is no cosmological horizon.
Indeed they can imply the existence of infinite numbers of
causally disconnected regions in the future, which were all
correlated in the past.  In the landscape scenario for string
theory \cite{kkltsuss} there are non-accelerating FRW cosmologies
in many of these regions, which have different degrees of freedom
and different space-time dimensions.   If one tries to think of
these models in terms of an S-matrix\footnote{This proposal is due
to L. Susskind.}, then the S-matrix for a given type of asymptotic
region is non-unitary.  Similarly \cite{crapsnjews} there are
treatments of cosmology in tree level string theory which seem to
imply multiple asymptotic regions, with the S matrix for any one
region violating unitarity.   If there are valid quantum theories
of this type, then one might equally well contemplate a similar
treatment of cosmologies with a number of e-foldings violating our
bound.   We will argue however that the dS region is unlikely to
be stable in such models, so our bound does not apply.

In this paper we will argue that the limited entropy that can fit
into asymptotic de Sitter space puts an upper bound on the amount
of inflation. On the other hand, there is a minimum number of
e-foldings required in order to reconcile the isotropy of the
microwave background on large scales with causality \footnote{For
an alternative, see work by Banks and Fischler on "holographic
cosmology" \cite{holcos}\cite{nextholo}} . We will see that these
numbers are rather close.

It is quite remarkable that a small cosmological constant,
seemingly irrelevant in magnitude when compared to the energy
density during inflation, has such an important impact. The
essential ingredient is that because of the UV-IR connection,
entropy requires storage space. The existence of a small
cosmological constant restricts the available storage space.

In the first section we present the details of how the bound on
the number of e-foldings is obtained. This will be followed by a
section where this bound is discussed in a different "holographic
gauge". We end by offering some conclusions.

\section{\bf The Number of e-Foldings is Bounded}

We begin by making a precise statement of the physical assumptions
that go into the derivation of our bound.   We first assume the
standard picture of inflationary cosmology, that is, that
inflation produces a large volume of space, of order $e^{3N_e}
\Lambda_I^{-{3\over 2}}$ in which all the degrees of freedom are
well described by local field theory, and are correlated with each
other. This system has a minimal entropy, which is needed to
describe the fluctuations in the energy density.    We will study
models in which the classical gravitational equations predict that
long after the inflationary period the universe enters an
asymptotically dS phase, with cosmological constant $\lambda$. We
then assume\cite{finitestates}, that the asymptotically static dS
observer has a complete, unitary description of all the physics
that she can measure\footnote{This means that her density matrix
is pure, if she makes all possible observations of the states on
her horizon.}, using a Hilbert space of finite dimension,
determined by $\lambda$. If the entropy needed to describe the
large inflationary patch exceeds the logarithm of the dimension of
this Hilbert space, then we have a contradiction. Either we must
give up the assumption of unitary evolution in causal patches, or
we must conclude that quantum fluctuations in the model, destroy
its classical asymptotic dS geometry.

Next, we briefly review the physics that leads to the bound on the
entropy that a fluid described by an equation of state $p =
\kappa\rho$ can encode. This physics was discussed in a recent
paper \cite{flp}. For the case of radiation, a hand-waving
derivation of the bound was also presented in \cite{tom}.

For any such fluid contained in a finite cavity of radius $R$, it
was shown in \cite {flp} that there is a threshold on the amount
of entropy stored in the fluid, beyond which black holes are
formed. The entropy $S_0$ at the verge of black hole formation
scales like

\eqn{S0}{S_0 \sim R^{3 - \frac{2}{1 + \kappa}}}

If we now imagine that the cavity is the region that is causally
accessible to the local observer living in asymptotic de Sitter
space-time with cosmological constant $\lambda$, then the length
scale in the previous equation $R$ is replaced by
${\lambda}^{-1/2}$.

This implies an upper bound on the entropy that scales
parametrically with $\lambda$ as follows

\eqn{S0lambda}{S_0 \sim {\lambda}^{-\frac{1 + 3\kappa}{2(1 +
\kappa)}}}

We will consider the  bound in the context of a
Lemaitre-Friedmann-Robertson-Walker (LFRW) cosmology where the
geometry can be described by the metric

\eqn{LFRW}{ds2 = dt^2 - a^2(t)(dr^2 +r^2d\Omega)}  where we choose a
spatially flat universe since we are interested in using the upper
bound on the entropy  at the exit from inflation.

The entropy $S_0$ carried by the dominant fluid in such a universe
can then be rewritten, using equilibrium thermodynamics,  as

\eqn{S0FRW}{S_0 = {{a_0}^3 {\rho}_0}^\frac{1}{1 + \kappa}}

The thermodynamical relation\footnote{ {In principle there is a
multiplicative constant of order one (in Planck units) in this
relation (in $1 + 1$ CFT it is related to $c$). It would weaken
our bound on the number of e-foldings only if the constant were
much less than one.  In CFT this constant is bounded from below by
a number of order one.  It affects the bound on $N_e$ by an
additive constant $\sim {\rm ln} c$.  For $\kappa = 1$ this
constant reflects the typical distance in Schwarzschild radii
between black holes in the black hole fluid, and the right
equation of state is only achieved when it is of order one.  For
non-relativistic matter of mass $M$ the coefficient is of order
${{M_P} \over {M}}$ and could increase the bound on $N_e$ by $\sim
10$, if the Compton wavelength is inside the Hubble radius. This
is still stronger than the bounds for stiffer fluids. } } that is
used, relates the energy density $\rho$ to the entropy density
$\sigma$

\eqn{thermo}{ \sigma \sim {\rho}^{\frac{1}{1+ \kappa}}}

In what follows, we will apply these ideas in the context of
inflation. Our approach is to first express the entropy according
to equation (4.2), where $a_0$ and ${\rho}_0$ are evaluated at the
exit of inflation. At the exit of inflation it is reasonable to
take for the energy density ${\rho}_0$  a value representative of
the energy density during inflation.

\eqn{rho}{{\rho}_0 \sim \Lambda_I} where $\Lambda_I$ is the value
of the energy density during inflation.

For the scale factor $a_0$ at the exit of inflation, we will take
the apparent horizon during inflation multiplied by the number of
e-foldings, $N_e$.

\eqn{a_0}{a_0 \sim {\Lambda_I}^{-1/2} e^{N_e}}

Substituting the expressions for $a_0$ and ${\rho}_0$ into
equation (2.4) leads, using equation (2.2) to the upper bound on
the number of e-foldings, $N_e$

\eqn{Ne}{N_e < \frac{1 + 3\kappa}{6(1 + \kappa)} log
\frac{{\Lambda}_I}{\lambda}}

Let us recapitulate what the physics involved in obtaining this
bound is. At the end of inflation, the energy density in the
universe becomes dominated by a fluid with an equation of state $p
= \kappa\rho$. The upper bound on the number of e-foldings for a
given $\kappa$ ,  is the threshold beyond which black holes form
and some of the excess entropy gets imprinted onto the asymptotic
de Sitter horizon. The biggest value for the number of e-foldings
occurs for the stiffest equation of state, $\kappa =1$.

\eqn{kappa1}{ N_e < \frac{1}{3}log \frac{{\Lambda}_I}{\lambda}}

This case is one that corresponds to a universe filled with black
holes \cite{holcos}\cite{blackcrunch}. For illustrative purpose,
we will estimate the value of $N_e$ for the case where after
inflation ends, the energy  is dominated by a $\kappa = 1$ fluid.
We will assume a value for $\Lambda_I \sim (10^{16} GeV)4$, which
is consistent with having not observed yet a background of
gravitons. We obtain the upper bound  on $N_e$

\eqn{Ne}{N_e \sim 85} where we took $\lambda$ to be of
$O({(10^{-3} eV)}^4)$.

For the sake of comparison, the case $\kappa = 1/3$ yields with
the same value for $\Lambda_I$

\eqn{kappa1/3}{N_e \sim 65}

This value for the maximum number of e-foldings is close to the
value necessary to solve the "horizon problem". It is interesting
to note that a small value for the number of e-foldings may have
observable implications for the low values of l in the spectrum of
fluctuations of the microwave background.

At this point, a field theorist may be puzzled because he can
write models of inflation where the inflaton is moving so slowly
that the number of e-foldings becomes enormous, that nonetheless
have solutions which appear to asymptote to a dS universe with
arbitrary value of the cosmological constant . The next section is
devoted to answering the legitimate concerns of such a field
theorist.

\section {The View from a Different "Holographic Gauge"}

The work of Bousso \cite{raph} shows that there are many different
ways to project the information in a space-time onto a collection
of holographic screens.  Different coordinate systems often
suggest different natural screens.  Thus, for example, in the
presence of a black hole, an external observer finds it natural to
project the information in the hole onto the horizon, at least
until the black hole begins to evaporate significantly.  An
infalling observer chooses a different sets of screens. Similarly,
in dS space, each static observer uses his cosmological horizon
for a screen. In \cite{nightmare} and \cite{davis} we suggested
that for certain states of the system there is an alternative
description, which approximates quantum field theory in global
coordinates better and better as the cosmological constant goes to
zero.   The states in question have only black holes which are
much smaller than the cosmological horizon, and the state in each
horizon volume is well described by local field theory.   The
holographic screens are chosen to be those of small causal
diamonds distributed throughout the volume of dS space-time.

The motivation for this suggestion was the observation that such
field theoretic states have an entropy per horizon volume of order
$R^{3\over 2}$, while the total dS entropy scales like $R2$, where
$R$ is the dS radius.  This indicates the possibility of a
description of field theoretic states by of order $R^{1\over 2}$
commuting copies of the field theory degrees of freedom in a
single horizon.

In quantum field theory in curved space time, dS space is, at late
times, an exponentially expanding sphere.  The number of copies of
a given horizon volume grows to infinity as the global time goes
to infinity.  Even if we impose a UV cutoff on the system, the
number of field theoretic degrees of freedom appears to go to
infinity with the global time.  The counting of degrees of freedom
above suggests that this picture is not valid for finite
$\lambda$, but requires an IR cutoff when the entropy implied by
the field theory picture exceeds the holographic bound in dS
space.   Field theory states with more entropy will back-react on
the geometry and,  when the back reaction is taken into account,
the system will no longer be asymptotically dS.   As we noted
above, these solutions with back reaction do not refer to the
quantum theory of asymptotically dS space.

This new holographic gauge allows us to understand why the
conventional treatment of inflation in a universe which is
asymptotically matter or radiation dominated is compatible with
the holographic principle.   This is true no matter how many
e-foldings there are.   We simply use the global gauge for the
inflationary phase of expansion.  Conventional restrictions on
inflationary models ensure that the states of the system are well
described by field theory.   Eventually, when we enter the
non-accelerating phase, the particle horizon can grow indefinitely
and there is no contradiction with the large entropy implied by
the field theoretic treatment of the inflationary stage.

However, if we truly have an asymptotically dS future, with
cosmological constant $\lambda$, there will be such a
contradiction.  Let us consider the case of radiation. As in the
previous section the entropy implied by conventional field theory
(which we can now understand as a holographic entropy in the
global gauge) is \eqn{infent}{S_{inf} =  e^{3N_e}
{\Lambda_I}^{-3/4}}. We can also view this from the global
holographic gauge of the asymptotic dS universe, with the true
cosmological constant.   Since the inflationary model has been
tuned so that horizon scale black holes do not form within the
present cosmological horizon, the state of the system qualifies as
a field theoretic state over scales larger than the horizon.
However, we have claimed that quantum gravity will put an IR
cutoff on this picture, determined by requiring that the volume of
space over which the global gauge is valid has field theoretic
entropy less than or equal to the total dS entropy .   This is a
refined version of the bound on the number of e-foldings announced
in the previous section.   We believe that the numerical value of
the bound obtained by this method, is more accurate than the
estimates in the first section.   Its form is

\eqn{newbd}{N_e \leq  {1\over 4} {\rm ln (\Lambda_I / \lambda)} +
{1\over 12} {\rm ln (M_P4 /\lambda )}} This gives $N_e \sim 88$,
for $\Lambda_I$ of order ${(10^{16}  GeV)}^4$.

It is easy to write down models of inflation with an arbitrarily
large number of e-foldings, which appear to asymptote to a dS
universe with fixed cosmological constant independent of $N_e$.
Consider for example a canonical chaotic inflation model
\cite{lindebook}, a single scalar, with potential \eqn{chaotic}{V
= g \phi4 + \lambda}.  To fit the amplitude of density
fluctuations observed in the CMB, $g$ is chosen to be $ g \sim
10^{-15}$.   The slow roll condition and the condition that the
energy density be less than the Planck scale are satisfied for a
range of $\phi / M_P$ of order $10^3$, and the number of
e-foldings is of order $10^3$. It is hard to understand how such a
fine tuned model could emerge from a fundamental theory like
string theory. Nonetheless, one would like to understand whether
such a model is inconsistent.

The spectrum of density fluctuations in models of this type is not
scale invariant over the whole range of scales which are generated
during inflation.   The parameter $g$ is tuned so that
${\delta\rho\over\rho}\sim 10^{-5}$, the experimental number,
during the last $25 $ or so e-foldings of inflation (which
generate fluctuations on the scales observed in the CMB).
However, during most of the inflationary history, and thus on most
scales larger than our horizon within the current inflationary
patch, the amplitude of fluctuations is much larger. This leads to
the following paradoxical situation.   We generally imagine that
fluctuations on scales much larger than our horizon scale cannot
effect us.  However, in inflationary models, the true particle
horizon is at least as large as the inflationary patch. Otherwise
field theory would not be able to generate correlations between
fluctuations on these large scales (assuming no correlations in
the initial state).   Although they will not affect us for a long
time, these large amplitude, large scale fluctuations do affect
the answer to the question of whether the universe is
asymptotically dS. We conjecture that these large amplitude
``superhorizon'' fluctuations can gravitationally collapse, and
our horizon volume will in fact be contained in a huge black hole.
An observer inside a black hole can remain oblivious to that fact
for times of order the Schwarzschild radius of the hole.   Since,
in the present situation, this radius is much larger than our
current horizon, there is no real contradiction with conventional
predictions of the model for experiments done today.   We are
simply saying that, in models of this sort, at some time in the
distant future, the currently observed accelerated expansion will
be replaced by a Big Crunch. Thus, the hypothesis that a model of
this type asymptotes to a dS universe is false, but it may still
give rise to a period of accelerated expansion from the point of
view of local observers. We emphasize that this is a conjectural
resolution of the contradiction posed by classical solutions with
many e-foldings and dS asymptotics.   Our bound is based on
quantum mechanical reasoning and cannot be evaded without giving
up one of the assumptions we have made.   We suggest that, in
these models, it is the assumption of asymptotic dS space which
must be relaxed, and are proposing a tentative mechanism for
changing the asymptotic behavior.

The global structure of inflationary models has been the subject
of intensive investigation \cite{lindeetal}, with results that the
present authors find confusing.  The standard field theory
arguments lead to a picture of eternal or self-reproducing
inflation.  We are not convinced that, in the absence of a
completely understood theory of quantum gravity, we are in a
position to understand these issues, but they do appear to imply
violations of unitarity for observers in causal patches. This
would remove the basic assumption underlying our bound.  However,
we think it is very unlikely that a dS observer would survive
forever in such a universe.

The question of finding global time slices in a self-reproducing
universe is a vexing one, but if the model is to be fit into the
conventional framework of quantum mechanics, it must have a
unitary time evolution operator $U(t, t_0)$.  The Hilbert space of
the asymptotically dS local observer is a finite dimensional
sub-factor of the vastly larger (perhaps infinite dimensional)
Hilbert space of the self-reproducing universe.  If our bound is
violated, then at the end of inflation, this factor is correlated
with a much larger sub-factor of the full space.    We know of no
example in quantum mechanics of a similar situation where the
small system does not decay and become mixed with the larger one.
Sometimes the decay proceeds by tunnelling, and takes a long time,
but it always occurs.   Indeed, in all semi-classical contexts in
which a dS minimum can interact with a system with a larger number
of degrees of freedom, it decays.

Finally, we want to mention an alternative to our bound which
seems to be what is assumed by various colleagues who have raised
objections to it \cite{lmb}.  Models with a number of e-foldings
violating our bound do not violate the covariant entropy bound.
Nonetheless, no observer in such a universe can access more than a
finite amount of information, given by the dS entropy.  This is
taken to suggest the existence of a quantum model, with a finite
dimensional Hilbert space which describes such universes.   How
can this be compatible with the evolution of the restricted state
vector of the causal patch from a pure to a mixed state?  The
suggestion is that this is an artifact of the field theory
approximation.   The real state of this system is pure, but
approximates the mixed state well.  It is surely true that there
are states like this in the causal patch Hilbert space.   What is
less than obvious is what the dynamics might be in this space,
which naturally evolves the system into this state.  To find it
one would have to construct the hypothetical quantum gravity model
which describes this space-time.  We can only admire the
confidence with which this opinion is stated, but we cannot share
it.

\section{Conclusions}

We have argued that a universe which is truly asymptotically dS is
only consistent with inflationary models which satisfy the bound
on the number of e-foldings.  However, our explanation of how this
is consistent with lagrangian models of inflation which can
incorporate an arbitrarily large number of e-foldings and an
arbitrary cosmological constant, implies that it might be
difficult to distinguish the two scenarios by observation. Namely,
we have argued that such models do not truly give rise to
asymptotically dS universes, but that the Big Crunch or other
instability which ends the dS era, might occur only after many
times the current age of the universe.

In this sense, models with a large number of e-foldings and a
positive cosmological constant are similar to quintessence models
like those of \cite{albrecht}, in which the current phase of
acceleration is modelled by something like old inflation: a
temporary sojourn in a meta-stable minimum with positive
cosmological constant.  This is followed by tunnelling out to a
region of rolling scalar field with no acceleration.   Such models
might arise naturally if string theory truly has a discretuum of
meta-stable dS vacua \cite{bpetal}

Models of this type raise the spectre of non-falsifiability. One
could imagine that no experiment in the foreseeable future could
distinguish between them and a model with a true asymptotic dS
space.

We can imagine two ways out of this uncomfortable situation. If
the hypothesis of Cosmological SUSY Breaking \cite{tbfolly} is
correct, then radiative corrections to the long distance effective
lagrangian may depend on the physics at large space-like distance,
in a significant way .  These virtual effects might then be able
to distinguish between different models.

A more exciting possibility is raised by the low $L$ anomalies in
the WMAP data.\footnote{This was emphasized to one of us (TB) at
the Middle East String Theory Conference in Crete \cite{sarkar}.}.
The WMAP data \cite{wmap} shows weak evidence of a problem for the
theoretical predictions of inflationary models for low values of
angular momentum.  Theoretical calculations of the amplitudes of
the fluctuations at fixed and small spherical harmonic, depends
somewhat on the theoretical predictions outside our current
horizon \cite{zeldov}. By introducing a break in the spectrum at
the horizon scale,  one can fix the ``problem". Indeed, in
Sarkar's model one can use the extra freedom to obtain a fit to
the data without a cosmological constant.   While we are (for
religious reasons) not particularly sympathetic to the latter
attempt, we would like to suggest that this aspect of the data may
eventually be viewed as evidence that inflationary models which
predict a scale invariant spectrum of fluctuations far outside the
current horizon are ruled out.   Our bound on the number of
e-folds would then suggest that a true asymptotic dS behavior was
the most likely explanation of this.

Inflationary models with small numbers of e-foldings exist.  They
are models with potentials of the form

\eqn{potential}{V = \frac{M^6} {M_P^2}\  U(\phi /{M_P} )} with $M$
of order the GUT scale.   They arise naturally in brane world
models derived from string theory, which use a mild difference
between the Kaluza-Klein scale and the higher dimensional Planck
scale to explain the ratio between the scales of gravity and
coupling unification.   The natural prediction of models of this
type is $O (1)$ e-folds.   One might imagine mild fine tuning of
parameters, or perhaps the sort of dynamical friction in
multi-field models that was discussed in \cite{BBMSS}, could
increase this to $10-100$, but more e-folds than that are
implausible.

Alternatively, one could turn to holographic cosmology
\cite{holcos}.  The current version of these models
\cite{nextholo} predicts a scale invariant spectrum, but only over a very
limited range of scales.   The range depends on parameters we can not yet
calculate but the largest scale is very unlikely to exceed the present
horizon by very much.

\acknowledgments We would like to thank A. Albrecht, E. di Napoli,
C. Krishnan, M. Zanic and especially A. Loewy, S. Paban and S.
Sarkar for useful discussions.  We thank R. Bousso, D.Lowe and
D.Marolf for discussions which led to the latest revision of this
work.

The research of W. Fischler is based upon work supported by the
National Science Foundation under Grant No. 0071512. The research
of T. Banks was supported in part by DOE grant number
DE-FG03-92ER40689.



\newpage
\begin{thebibliography}{19}






\bibitem{deBernardis:2001xk}
P.~de Bernardis {\it et al.}, {\it Multiple peaks in the angular
power spectrum of the cosmic microwave background: Significance
and consequences for cosmology}, Astrophys.\ J.\  {\bf 564}, 559
(2002) [arXiv:astro-ph/0105296].

\bibitem{Balbi:2000tg}
A.~Balbi {\it et al.}, { \it Constraints on cosmological parameters
from MAXIMA-1}, Astrophys.\ J.\  {\bf 545}, L1 (2000)
[Erratum-ibid.\  {\bf 558}, L145 (2001)] [arXiv:astro-ph/0005124].

\bibitem{Stompor:2001xf}
R.~Stompor {\it et al.}, { \it Cosmological implications of the
MAXIMA-I high resolution Cosmic Microwave Background anisotropy
measurement}, Astrophys.\ J.\  {\bf 561}, L7 (2001)
[arXiv:astro-ph/0105062].

\bibitem{Abroe:2001de}
M.~E.~Abroe {\it et al.}, {\it Frequentist Estimation of Cosmological
Parameters from the MAXIMA-1 Cosmic Microwave Background
Anisotropy Data}, Mon.\ Not.\ Roy.\ Astron.\ Soc.\  {\bf 334}, 11
(2002) [arXiv:astro-ph/0111010].

\bibitem{Pryke:2001yz}
C.~Pryke, N.~W.~Halverson, E.~M.~Leitch, J.~Kovac,
J.~E.~Carlstrom, W.~L.~Holzapfel and M.~Dragovan, {\it Cosmological
Parameter Extraction from the First Season of Observations with
DASI}, Astrophys.\ J.\  {\bf 568}, 46 (2002)
[arXiv:astro-ph/0104490].

\bibitem{Spergel:2003cb}
D.~N.~Spergel {\it et al.}, {\it First Year Wilkinson Microwave
Anisotropy Probe (WMAP) Observations: Determination of
Cosmological Parameters}, astro-ph/0302209.

\bibitem{Perlmutter:1997ds}
S.~Perlmutter {\it et al.}  [Supernova Cosmology Project
Collaboration], {\it Measurements of the Cosmological Parameters
$\Omega$ and $\Lambda$ from the First 7 Supernovae at z $\geq$
0.35}, Astrophys.\ J.\ {\bf 483}, 565 (1997) [astro-ph/9608192].

\bibitem{Riess:1998cb}
A.~G.~Riess {\it et al.}  [Supernova Search Team Collaboration],
{\it Observational Evidence from Supernovae for an Accelerating
Universe and a Cosmological Constant}, Astron.\ J.\  {\bf 116},
1009 (1998) [arXiv:astro-ph/9805201].

\bibitem{Primack:2002th}
J.~R.~Primack, {\it Status of cold dark matter cosmology},
arXiv:astro-ph/0205391.

\bibitem{lennybf} See for example: L.~Dyson, J.~Lindesay,
L.~Susskind, {\it Is there really a de Sitter/CFT duality}, JHEP
{\bf0208}, 045, (2002), hep-th/0202163; T.~Banks, W.~Fischler,
{\it M-theory observables for cosmological spacetimes},
hep-th/0102077.

\bibitem{flp} W.~Fischler, A.~Loewy, S.~Paban, {\it The Entropy of the Microwave Background and the Acceleration
of the Universe}, hep-th/0307031.

\bibitem{tom} T.~Banks, {\it Some Thoughts on the Quantum
Theory of de Sitter Space}, astro-ph/0305037; {\it A critique of
pure string theory: Heterodox opinions of diverse dimensions},
hep-th/0306074.


 \bibitem{finitestates} T.~Banks, {\it Cosmological breaking of supersymmetry? or little lambda goes back
to the future 2}, hep-th/0007146; W.~Fischler, {\it Taking de
Sitter seriously.}, Talk given at {\it Role of scaling laws in
physics and biology (celebrating the 60th birthday of Geoffrey
West)}, Santa Fe, Dec., 2000.

\bibitem{GH}
G.~W.~Gibbons and S.~W.~Hawking, {\it Cosmological Event Horizons,
Thermodynamics, And Particle Creation} Phys.\ Rev.\ D {\bf 15},
2738 (1977).

\bibitem{rbousso} R.~Bousso, {\it Positive vacuum energy and the
N-bound}, JHEP {\bf0011}, 038, (2000), hep-th/0010252; see also
R.~Bousso, O.~DeWolfe, R.~Myers, {\it Unbounded entropy in
spacetimes with positive cosmological constant}, hep-th/0205080.

\bibitem{holcos} T.~Banks, W.~Fischler,
{\it An Holographic Cosmology}, hep-th/0111142

\bibitem{nextholo} T.~Banks, W.~Fischler, {\it Holographic Cosmology
3.0}, hep-th/0310288

\bibitem{blackcrunch} T.~Banks, W.~Fischler, {\it Black Crunch}, hep-th/0212113.

\bibitem{raph}  R.~Bousso, {\it Holography in general
space-times,} JHEP { \bf 9906}, 028 (1999) hep-th9906022

R.~Bousso, {\it The Holographic Principle} Rev.Mod.Phys. {\bf 74}: 825
(2002), hep-th/0203101

\bibitem{nightmare}  T.~Banks, W.~Fischler, S.~Paban,
{\it Recurrent Nightmares?: Measurement Theory in de Sitter
Space}, JHEP {\bf 0212}, 062 (2002) hep-th/0210160

\bibitem{davis}  T.~Banks, {\it Some thoughts on the quantum theory of de Sitter space},
astro-ph/0305037

\bibitem{lindebook} For a detailed discussion on chaotic inflation see: A.D. Linde, {\it Particle
Physics and Inflationary Cosmology}   (Harwood Academic
Publishers, Chur, Switzerland 1990).

\bibitem{lindeetal} A.~D.~Linde, D.~A.~Linde and A.~Mezhlumian,
{\it From the Big Bang theory to the theory of a stationary
universe}, Phys.\ Rev.\ D {\bf 49}, 1783 (1994)
[arXiv:gr-qc/9306035].

\bibitem{albrecht} A.~Albrecht, C.~Skordis {\it Phenomenology of a realistic accelerating universe using
only Planck scale physics}, Phys. Rev. Lett. {\bf 84}, 2076 (2000),
astro-ph/9908085

\bibitem{bpetal} R.~Bousso, J.~Polchinski,{\it Quantization of four form fluxes and dynamical
neutralization of the cosmological constant} JHEP {\bf 0006}, 006
(2000), hep-th/0004134

J.L.~Feng, J.~March-Russell, S.~Sethi, F.~Wilczek,{\it Saltatory
relaxation of the cosmological constant}, Nucl.Phys.{\bf B602},
307 (2001), hep-th/0005276

S.B.~Giddings, S.~Kachru, J.~Polchinski, {\it Hierarchies from
fluxes in string compactifications}, Phys.Rev.{\bf D66}, 106006
(2002), hep-th/0105097

S.~Kachru, R.~Kallosh, A.~Linde, S.P.~Trivedi,{\it de Sitter vacua
in string theory}, hep-th 0301420

L.~Susskind, {\it The anthropic landscape of string theory},
hep-th/0302219

\bibitem{tbfolly} T.~Banks, {\it Cosmological breaking of supersymmetry? or little lambda goes back
to the future 2}, hep-th/0007146

\bibitem{sarkar} S.~Sarkar, {\it Talk at the Middle East
String Theory Conference, Orthodox Academy of Crete, Kolymbari
Crete, June 20-27}

A.~Blanchard, M.~Douspis, M.~Rowan-Robinson, S.~Sarkar, {\it An
alternative to the cosmological concordance model}
astro-ph/0304237, to appear in Astronomy `I\&' Astrophysics

R.~Scranton et al. (SDSS Colaboration), {\it Physical evidence for
dark energy} astro-ph/0307335

G.~Efstathiou, {\it the statistical significance of the low CMB
multipoles} astro-ph/0306431

G.~Efstathiou, {\it Is the low CMB quadrupole a signature of
spatial curvature} astro-ph/0303127

J.M.~Cline, P.~Crotty, J.~Lesgourgues, {\it Does the small CMB
quadrupole moment suggest new physics} astro-ph/0304558

B.~Feng, X.~Zhang, {\it Double inflation and the low CMB
quadrupole}, astro-ph/0305020.

A.~de Oliveira-Costa, M.~Tegmark, M.~Zaldarriaga, A.~Hamilton,
{\it The significance of the largest scale CMB fluctuations in
WMAP} astro-ph/0307282

\bibitem{zeldov} L.P.~Grishuk, Ya.B.~Zeldovich, Sov. Astron. {\bf 22} (1978) 125

For a more recent discussion see: D.H.~Lyth, A.~Woszczyna, {\it
Large scale perturbations in the open universe} Phys. Rev.{\bf
52},3338 (1995), astro-ph/9501044


\bibitem{BBMSS}, T.~Banks, M.~Berkooz, G.~Moore, S.~Shenker, P.~Steinhardt, {\it Modular cosmology},
Phys. Rev. {\bf D52}, 705 (1995), hep-th/9501053

\bibitem{wmap}, H.V.~Peiris, E.~Komatsu, L.~Verde, D.N.~Spergel, C.L.~Bennett, M.~Halpern, A.~Kogut,
G.~Hinshaw, N.~Jarosik, M.~Limon, S.S.~Meyer, {\it First year
Wilkinson anisotropy probe (WMAP) observations: implications for
inflation}, astro-ph/0302225

\bibitem{mcosmo}  T.~Banks, W.~Fischler,
{\it M-theory observables for cosmological spacetimes},
hep-th/0102077.

\bibitem{tHsuss} 
G.~t'Hooft, {\it Dimensional reduction in quantum gravity}, gr-qc/9310026

L.~Susskind, {\it The world as a hologram}, J.Math.Phys. 36, 6377 (1995),
hep-th/9409089

\bibitem{kkltsuss}  S.~Kachru, R.~Kallosh, A.~Linde, S.P.~Trivedi,{\it de Sitter vacua
in string theory}, hep-th 0301420

L.~Susskind, {\it The anthropic landscape of string theory},
hep-th/0302219

\bibitem{crapsnjews} S.~Elitzur, A.~Giveon, D.~Kutasov,
E.~Rabinovici,{\it From big bang to big crunch and beyond}, JHEP 0206,
017 (2002), hep-th/0204189 

 B.~Craps, D.~Kutasov, G.~Rajesh, {\it String propagation in
the presence of cosmological singularities} JHEP 0206, 053 (2002),
hep-th/0205101

M.~Berkooz, B.~Craps, D.~Kutasov, G.~Rajesh,{\it Comments on cosmological
singularities in string theory}, JHEP 0303, 031 (2003), hep-th/0212215

\bibitem{lmb} D.~Lowe, D.~Marolf, {\it Holography and eternal inflation},
hep-th/0402162

R.~Bousso, private communication



\end{thebibliography}
\end{document}